\documentclass[review,12pt]{elsarticle}

\usepackage{lineno}
\modulolinenumbers[5]

\def\bff{{\mathbf{f}}}

\def\bm{{\mathbf{m}}}

\def\bu{{\mathbf{u}}}

\def\bx{{\mathbf{x}}}
\def\by{{\mathbf{y}}}
\def\bz{{\mathbf{z}}}
\def\b0{{\mathbf{0}}}

\def\bI{{\mathbf{I}}}

\def\bK{{\mathbf{K}}}

\def\bQ{{\mathbf{Q}}}

\def\bS{{\mathbf{S}}}

\def\bX{{\mathbf{X}}}

\def\bZ{{\mathbf{Z}}}

\def\btheta{{\boldsymbol{\theta}}}

\def\bmu{{\boldsymbol{\mu}}}

\def\bSigma{{\boldsymbol{\Sigma}}}

\def\rd{{\mathrm{d}}}

\newcommand{\p}{{\mathrm p}}
\newcommand{\q}{{\mathrm q}}

\def\R{{\mathbb{R}}}
\usepackage{color}

\usepackage{amsfonts}
\usepackage{amssymb}
\usepackage{amsbsy} 
\usepackage{amsmath}
\usepackage{graphicx}
\usepackage{epstopdf}
\usepackage{appendix}
\usepackage{amsthm}
\usepackage{bbm}
\usepackage{framed}
\usepackage{soul}

\newcommand{\tit}{Deep Gaussian Processes for Parameter Retrieval}

\usepackage{array}
\newcolumntype{L}[1]{>{\raggedright\let\newline\\\arraybackslash\hspace{0pt}}m{#1}}
\newcolumntype{C}[1]{>{\centering\let\newline\\\arraybackslash\hspace{0pt}}m{#1}}
\newcolumntype{R}[1]{>{\raggedleft\let\newline\\\arraybackslash\hspace{0pt}}m{#1}}

\usepackage{pgffor} 
\usepackage{float} 
\usepackage{wrapfig}
\usepackage{flushend}  

\usepackage{amsmath}
\usepackage{amssymb}
\usepackage{amsthm}
\usepackage{algorithm}
\usepackage{algorithmic}
\usepackage{placeins}
\usepackage{multirow}
\usepackage{color}
\usepackage{url}
\usepackage{cite}
\usepackage{wrapfig,lipsum,booktabs}
\usepackage[font=small,labelfont=bf]{caption}
\usepackage{graphicx}
\usepackage{xcolor}
\usepackage[all]{xy}
\usepackage{hyperref}
\usepackage{adjustbox}
\hypersetup{
    unicode=false,          
    pdftoolbar=true,        
    pdfmenubar=true,        
    pdffitwindow=false,     
    pdfstartview={FitH},    
    pdftitle={\tit},    
    pdfauthor={Adrian Perez-Suay and Camps-Valls},     
    pdfsubject={\tit},   
    pdfcreator={Adrian Perez-Suay and Camps-Valls},   
    pdfproducer={Adrian Perez-Suay and Camps-Valls}, 
    pdfkeywords={forward}{inverse}{modeling}{dependence}, 
    pdfnewwindow=true,      
    colorlinks=true,       
    linkcolor={blue},          
    citecolor=blue,        
    filecolor=blue,      
    urlcolor=blue           
}
\usepackage{pgffor} 

\usepackage{setspace}
\makeatletter
\def\ps@pprintTitle{%
  \let\@oddhead\@empty
  \let\@evenhead\@empty \textsuperscript{\textcopyright}Elsevier. Accepted for publication in ISPRS Journal of Photogrammetry and Remote Sensing. DOI 10.1016/j.isprsjprs.2020.04.014
  \let\@oddfoot\@empty
  \let\@evenfoot\@oddfoot
}
\makeatother

\begin{document}

\begin{frontmatter}

\title{Deep Gaussian Processes for Biogeophysical Parameter Retrieval and Model Inversion}

\author[uv]{Daniel Heestermans Svendsen\corref{mycorrespondingauthor}}
\cortext[mycorrespondingauthor]{Corresponding author}
\ead{daniel.svendsen@uv.es}
\author[ug]{Pablo Morales-\'Alvarez}
\ead{pablomorales@decsai.ugr.es}
\author[uv]{Ana Belen Ruescas}
\ead{ana.b.ruescas@uv.es}
\author[ug]{Rafael Molina}
\ead{rms@decsai.ugr.es}
\author[uv]{Gustau Camps-Valls}
\ead{gustau.camps@uv.es}

\address[uv]{Image Processing Lab (IPL), Universitat de Val\`encia, C/ Cat. Jos\'e Beltr\'an, 2. 46980 Paterna, Spain.}
\address[ug]{Department of Computer Science and Artificial Intelligence, University of Granada, 18010 Granada, Spain}

\begin{abstract}
Parameter retrieval and model inversion are key problems in remote sensing and Earth observation. Currently, different approximations exist: a direct, yet costly, inversion of radiative transfer models (RTMs); the statistical inversion with {\em in situ} data that often results in problems with extrapolation outside the study area; and the most widely adopted hybrid modeling by which statistical models, mostly nonlinear and non-parametric machine learning algorithms, are applied to invert RTM simulations. We will focus on the latter. Among the different existing algorithms, in the last decade kernel based methods, and Gaussian Processes (GPs) in particular, have provided useful and informative solutions to such RTM inversion problems. This is in large part due to the confidence intervals they provide, and their predictive accuracy. However, RTMs are very complex, highly nonlinear, and typically hierarchical models, so that very often a single (shallow) GP model cannot capture complex feature relations for inversion. This motivates the use of deeper hierarchical architectures, while still preserving the desirable properties of GPs. This paper introduces the use of deep Gaussian Processes (DGPs) for bio-geo-physical model inversion. Unlike shallow GP models, DGPs account for complicated (modular, hierarchical) processes, provide an efficient solution that scales well to big datasets, and improve prediction accuracy over their single layer counterpart. In the experimental section, we provide empirical evidence of performance for the estimation of surface temperature and dew point temperature from infrared sounding data, as well as for the prediction of chlorophyll content, inorganic suspended matter, and coloured dissolved matter from multispectral data acquired by the Sentinel-3 OLCI sensor. The presented methodology allows for more expressive forms of GPs in big remote sensing model inversion problems.
\end{abstract}

\end{frontmatter}



\section{Introduction} \label{sec:int}
Estimating variables and bio-geophysical parameters of interest from remote sensing images is a central problem in Earth observation~\citep{Liang08,rodgers00,CampsValls11mc}. 
This is usually addressed through a very challenging {\em model inversion problem}, which involves dealing with complex nonlinear input-output relations. 
In addition, very often, the goal is to invert {\em metamodels}, that is, combinations of submodels that are coupled together. In remote sensing, radiative transfer models (RTMs) describe the processes which occur at different scales (e.g. at leaf, canopy and atmospheric levels) with different complexities. The overall process is thus complicated, nonlinear and hierarchical, with different sources of uncertainty propagating through the system.

The inversion of such highly complex models has been attempted through several strategies. One standard approach consists on running a reasonable number of RTM simulations which generates the so called look-up tables (LUTs). Then, for a new input observation, one assigns the most similar parameter in the LUT. A second, more direct approach involves the direct physics-based inversion, which results in complex optimization problems. An alternative {\em hybrid approach} comes from the use of statistical approaches to perform the inversion using the LUT simulations.  
A review of approaches can be found in~\citep{verrelst12b,CampsValls11mc}. 
In recent years, the remote sensing community has turned to this type of statistical {\em hybrid} approaches for model inversion~\citep{CampsValls11mc}, {mainly because of efficiency, versatility and the interesting balance between its data driven and physics-aware nature~\citep{verrelst2015experimental}}. \\
Approximating arbitrary nonlinear functions from data is a solid field of machine learning where many successful methods are available.
Data-driven statistical learning algorithms have attained outstanding results in the estimation of climate variables and related geo-physical parameters at local and global scales~\citep{CampsValls09wiley,CampsValls11mc}. 
These algorithms avoid complicated assumptions and provide flexible non-parametric models that fit the observations using large heterogeneous data. The fact is that a plethora of regression algorithms have been used. There exist traditional models {such as random forests~\citep{Tramontana16bg,Jung17nature} and standard feed-forward neural networks~\citep{Blackwell05,Blackwell08,Camps-Valls20121759} as well as convolutional neural networks~\citep{malmgren2019statistical,ma2019deep}.}

In the last decade, more emphasis has been put on kernel methods in general~\citep{CampsValls09wiley,Rojo18dspkm}, and Gaussian Processes (GPs) in particular. There is a considerable amount of reasons for this.
Firstly, GPs constitute a probabilistic treatment of regression problems leading to an analytical expression for the {predictive uncertainty which is an attractive feature~\citep{Verrelst2013c,schneider2014evaluating}}. This also allows for effective error propagation from the inputs to the outputs as has recently been shown in~\citep{johnson2019accounting}. Furthermore, GPs are not pure black box models because, through the use and design of appropriate covariance functions, one can include prior knowledge about the signal characteristics (e.g. nonstationarity, heteroscedasticity, etc.). The covariance hyperparameters are learned (inferred) from data so that the model is interpretable. For instance, by using the automatic relevance determination (ARD) covariance function~\citep{verrelst2016spectral}, an automatic feature ranking can be derived from the trained model, thus leading to a explanatory model. These theoretical and practical advantages have recently translated to a wider adoption by the geoscience and remote sensing community in many applications and products, such as the spatialization of in-situ measurements and upscaling of carbon, energy and water fluxes~\citep{jung2017compensatory}. Gaussian Processes have provided very good results for retrieval in all Earth science domains, {be it land and vegetation parameter retrieval~\citep{Furfaro06,rivera2017hyperspectral,CampsValls18sciasi,CampsValls19nsr}}, ocean and water bodies modeling~\citep{Ruescas18rs,Sarkar19}, cryosphere ice sheet modeling and process emulation~\citep{Wernecke19}, or atmospheric parameter retrieval~\citep{CAMPS2012}.

Despite being successful in many different applications, standard GPs have two important shortcomings we want to highlight: 
\begin{itemize}
    \item {\em Computational cost.} A standard GP, which stores and uses all the data at once, exhibits a high computational cost. These GPs scale cubically with the number of data points when training, and quadratically when doing prediction.
    This hampers their adoption in applications which involve more than just a few thousand input points.
    \item {\em Expressiveness.} GPs are shallow models\footnote{
    It can be shown that, in the limit and under some mild assumptions, a GP corresponds to a single-hidden layer neural network with infinite neurons~\citep{neal1996priors}.}, so while accurate and flexible, their expressive power is limited when dealing with hierarchical structures. This is even worse due to the (ab)use of standard kernel functions like the exponentiated quadratic family (e.g. the RBF kernel is infinite-differentiable and tends to oversmooth functions).
\end{itemize}  

\begin{figure}[t!]
\centerline{
\includegraphics[width=11cm]{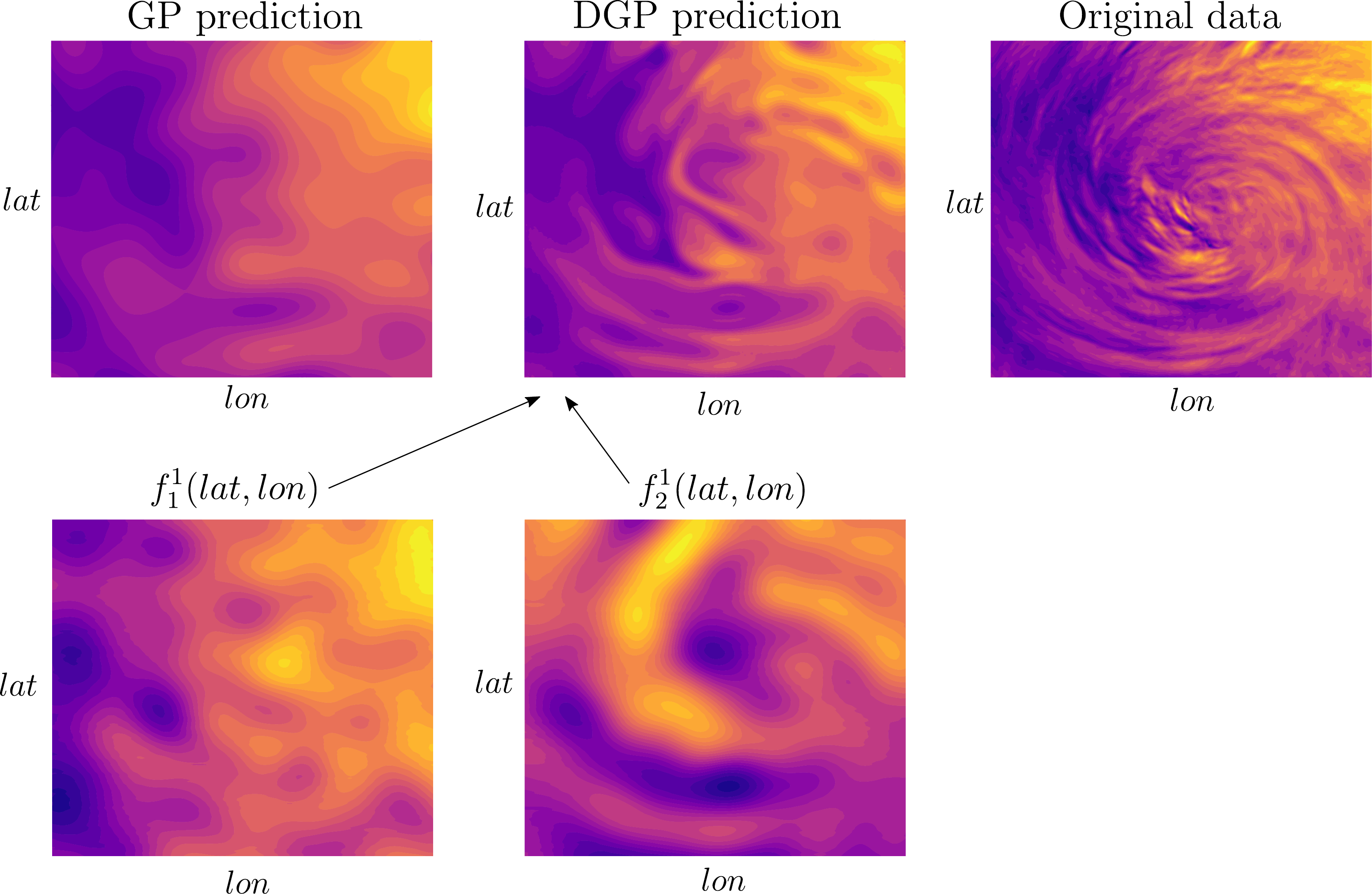}}

\caption{Example of shallow versus deep GPs. Modeling a hurricane field from the coordinates using $1000$ randomly selected training points in both cases. The GP prediction [top left] is too blurry and does not capture the whirl data structure (the scale of feature relations changes along the hurricane ridges). The DGP model [top middle] uses only two latent functions in its first layer. The first latent function $f_1$ captures lower frequencies [bottom left] --similarly to the GP map-- and the $f_2$ [bottom middle] focuses on the hurricane structure, while their combination leads to an overall predictive function [top middle] that better approximates the observation [top right].}
\label{fig:supertoy} 
\end{figure}

The first limitation is typically addressed through sparse GPs \citep{snelson2006sparse}, which have already been used in remote sensing applications \citep{morales2017remote}. 
In order to additionally tackle the second limitation, in this paper we introduce the use of Deep Gaussian Process (DGP) \citep{salimbeni_doubly_2017} to the field of remote sensing for the first time. A DGP is a cascaded and hierarchical model that captures more complex data structures, while still being able to scale well to millions of points.
 Our proposal is not incidental: the complexity of the processes involved in geosciences and remote sensing leads to highly hierarchical and modular models to be inverted. This calls for the application of the most innovative available techniques as shown in the following example. Fig.~\ref{fig:supertoy} compares the use of a standard GP and deep GP to model a hurricane structure.
It becomes clear that, unlike GPs, the DGP can cope with the whirl structure efficiently by combining different latent functions hierarchically. 

DGPs were originally introduced in~\citep{titsias_bayesian_2010,damianou2013deep}, and further analyzed in~\citep{damianou_deep_2015}.
In \citep{svendsen2018deep}, we outlined the potential use of DGPs for surface level dew point temperature retrieval from sounding data.
In this paper we extend that work in several ways: 
1) we focus the analysis on large scale remote sensing problems, aiming for a complete treatment of the two aforementioned standard GP shortcomings; 
2) provide a deeper formalization and more intuitive insight on the model for the practitioner; 
3) give more empirical evidences of performance in ocean and land parameter retrieval applications, and using different sensory data (optical sensors and microwave sounders); and 
4) assess accuracy and robustness to sample sizes and problem dimensionality versus both standard and sparse implementations of GPs.

In short, this work exposes the DGP methodology to the remote sensing community for the first time and for a wide range of applications.
The proposed DGP appears to be an excellent approach for model inversion.
Moreover, sticking to the GP framework is very convenient. GPs are based on a solid Bayesian formalism and inherit all properties of a probabilistic treatment: possibility to derive not only point-wise predictions but also confidence intervals, perform error quantification and uncertainty propagation easily, and optimize hyperparameters by log-likelihood maximization.

The remainder of the paper is organized as follows. Section~\ref{sec:theo} establishes notation, reviews the probabilistic modeling and inference of GP and sparse GP, and presents the deep GP model - mathematical details on modeling, inference, and prediction are provided in appendices \ref{sec:ap_FITC}, \ref{sec:appendix}, and \ref{sec:predictions}. Section~\ref{sec:exp} provides the experimental results. We illustrate performance in prediction of surface temperature and dew point temperature (related to relative humidity) from superspectral infrared sounding data~\citep{aires02,simeoni1997,huang1992}; as well as for the estimation of predicting chlorophyll content, inorganic suspended matter, and coloured dissolved organic matter from simulated multispectral data acquired by Sentinel-3 OLCI sensor. Finally, Section~\ref{sec:con} concludes the paper with summarizing remarks.

\section{Probabilistic Model and Inference} \label{sec:theo}

In this section we provide a brief and graphical introduction to modeling and inference for Gaussian Processes (GP) and Deep Gaussian Processes (DGP) in supervised regression problems. We explain and graphically show the hierarchical structure of DGPs, and also explain how both GPs and DGPs,  make use of sparse approximations to perform inference tasks. Mathematical details are deferred to appendices.

Gaussian Processes are non-parametric probabilistic state-of-the-art models for functions, and are successfully used in supervised learning.
In geostatistics, GPs for regression is usually referred to as \emph{kriging}.
The main strength of GPs is their accurate uncertainty quantification, which is a consequence of its sound Bayesian formulation, yielding well-calibrated predictions and confidence intervals \citep{Rasmussen2006, damianou_deep_2015}.

More specifically, for input-output data $\{(\bx_i,y_i)\in\R^d\times\R\}_{i=1}^n$, a GP models the underlying dependence with latent variables $\{f_i = f(\bx_i)\in\R\}_{i=1}^n$ that jointly follow a Gaussian distribution $\p(\bff)=\mathcal{N}\left(\bff|\mathbf{0},\bK\right)$.
The kernel matrix $\bK = (k(\bx_i,\bx_j))_{i,j}$ encodes the properties (e.g. smoothness) of the modeled functions.
The most popular standard kernel is the squared exponential one (or RBF), which is given by $k(\bx,\by)=\gamma\cdot\exp\left(-||\bx-\by||^2/(2\sigma^2)\right)$, with $\gamma$ (variance) and $\sigma$ (length-scale) called the kernel hyperparameters. 
Finally, in regression problems, the observation model of the outputs $y_i$ given the latent variables $f_i$ is usually defined by the Gaussian $\p(y_i|f_i,\rho^2)=\mathcal{N}(y_i|f_i,\rho^2)$. The variance $\rho^2$ is estimated during the training step, along with the kernel hyperparameters, by maximizing the marginal likelihood of the observed data.

Since the Gaussian prior $\p(\bff)$ is conjugate to the Gaussian observation model, one can integrate out $\bff$ and compute the marginal likelihood $\p(\by)$ and the posterior $\p(\bff|\by)$ in closed form (parameters are omitted for simplicity) \citep{Rasmussen2006}. However, this requires inverting the $n\times n$ matrix $(\bK+\rho^2 \bI)$, which scales cubically, $\mathcal{O}(n^3)$ where $n$ is the number of training data points. This constraint makes GP prohibitive for large scale applications, with $n=10^4$ usually being considered the practical limit \citep{morales2017}. Here, sparse GP approximations become the preferred pathway to scale the desirable properties of GPs to larger datasets \citep{snelson2006sparse, gpBigData, bauer2016, morales2017}, and they will be reviewed in Section \ref{sec:sparseGP}.
Interestingly, we will see that DGPs preserve the scalability of sparse GP approximations (while achieving a higher expressiveness).

\begin{figure}
    \centering
    \includegraphics[width=0.7\columnwidth]{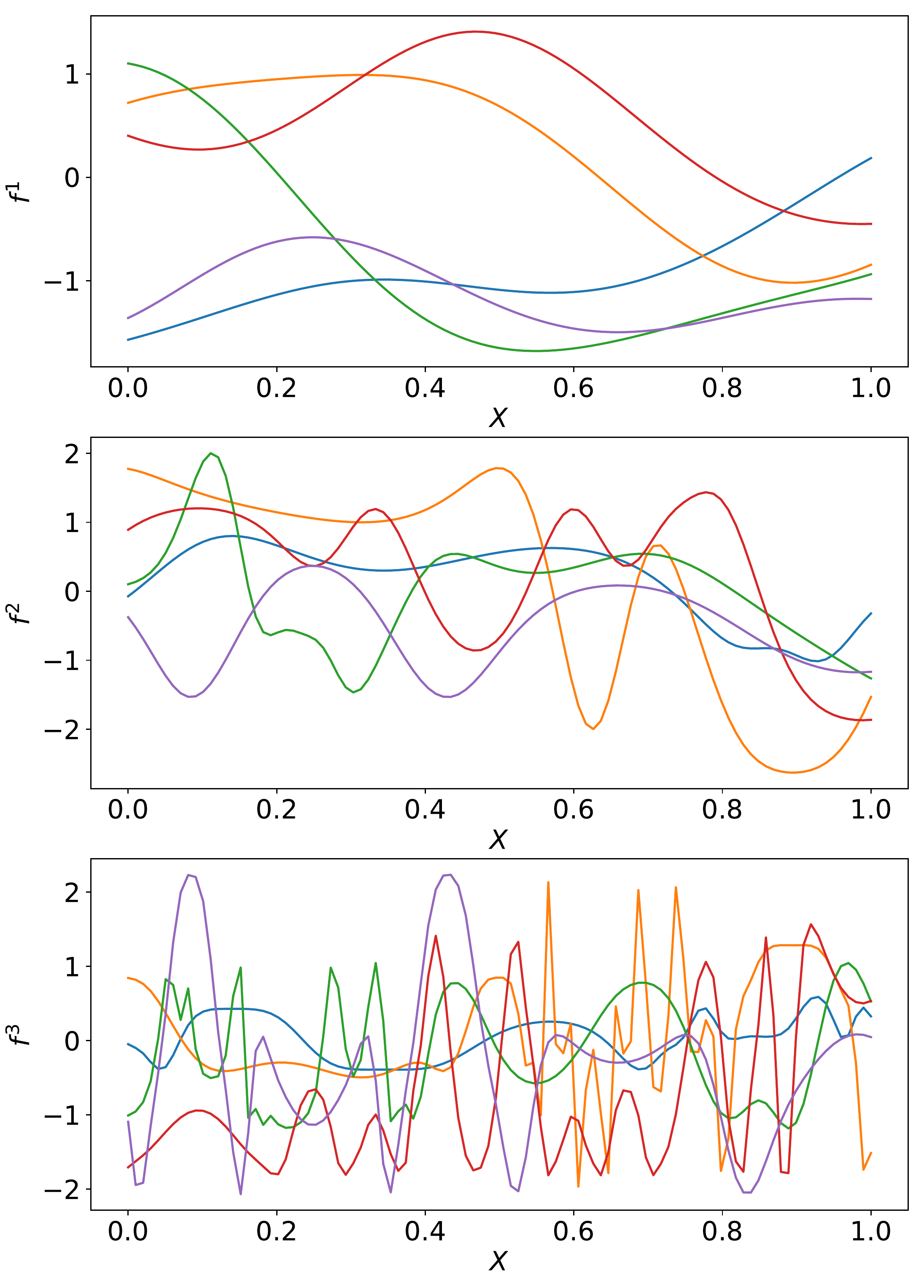}
    \caption{Five random samples from a 1-dimensional DGP with three layers and one hidden unit per layer. Each function sample uses the function of the same color in the previous plot as input, except the function samples of the top plot (L=1) which use the actual values of x as input.
    Every layer is endowed with a standard RBF kernel. 
    This produces very smooth functions in the first layer (i.e. a shallow GP, top plot). However, the concatenation of such simple GPs produces increasingly complex functions (middle and bottom plots, 2-layer and 3-layer DGPs respectively). In particular, notice that DGP-3 captures sophisticated patterns that combine flat regions with high-variability ones, which cannot be described by stationary kernels. These ideas are behind the superiority of DGPs in Fig. \ref{fig:supertoy}. 
    }
    \label{fig:DGP_samples}
\end{figure}

Additionally, GPs are limited by the expressiveness of the kernel function. Ideally, complex kernels could be tailored for different applications \citep{Rasmussen2006}. However, this is usually unfeasible in practice, as it requires a thorough application-specific knowledge. Moreover, it usually comes with a large amount of hyperparameters to estimate, which may cause overfitting. As a result, standard general-purpose kernels are normally considered in practice. Alternatively, DGPs allow for modeling very complex data through a hierarchy of GPs that only use simple kernels with few hyperparameters as building blocks (like the aforementioned RBF one, which will be used here). 
Fig.~\ref{fig:DGP_samples} provides an intuition on this, and DGPs will be introduced in Section \ref{sec:DGP}.

\subsection{Sparse GP approximations}\label{sec:sparseGP}

In the last years, many different sparse GP approximations have been introduced in order to cope with increasingly large datasets \citep{snelson2006sparse, gpBigData, bauer2016, morales2017}. 
Most of them resort to the notion of \emph{inducing points}, a reduced set of $m\ll n$ latent variables which the inference is based on. More specifically, these inducing points $\bu=(u_1,\dots,u_m)$ are GP realizations at the \emph{inducing locations} $\bZ=\{\bz_1,\dots,\bz_m\}\subset\R^d$, just like $\bff$ is at the inputs $\bX=\{\bx_1,\dots,\bx_n\}$.
All these sparse methods are grouped in two big categories, depending on where exactly the approximation takes place: in the model definition or in the inference procedure \citep{bauer2016}. Both types of sparse GP will be compared against the deep GP in the experiments.

In the first group, the \emph{Fully Independent Training Conditional} (FITC) \citep{snelson2006sparse} is the most popular approach. It uses the inducing points to approximate the GP model, and then marginalizes them out and perform exact inference. This yields a reduced $\mathcal{O}(nm^2)$ computational complexity (linear in the dataset size). 
Mathematical details for FITC are included in Appendix \ref{sec:ap_FITC}.

In the second group, the \emph{Scalable Variational Gaussian Process} (SVGP) \citep{gpBigData} is one of the most widespread methods.
It maintains the exact GP model, and uses the inducing points to introduce the approximation in the inference process through variational inference \citep{blei2017variational}.
The mathematical details are included in Appendix \ref{sec:appendix} (which is devoted to DGPs because, as we explain in next paragraph, SVGP is equivalent to DGP with one layer).
Since SVGP does not modify the original model, it is less prone to overfitting. However, if the posterior distribution is not well approximated within the variational scheme, its performance might become poorer. Therefore, both groups of methods are complementary, and in the machine learning community none of them is considered to consistently outperform the other \citep{bauer2016}. An advantage of SVGP over FITC is its factorization in mini-batches, which allows for even greater scalability. In this case, the computational cost is $\mathcal{O}(n_bm^2)$, with $n_b$ the mini-batch size.

\begin{figure}
    \centering
    \includegraphics[width=0.8\columnwidth]{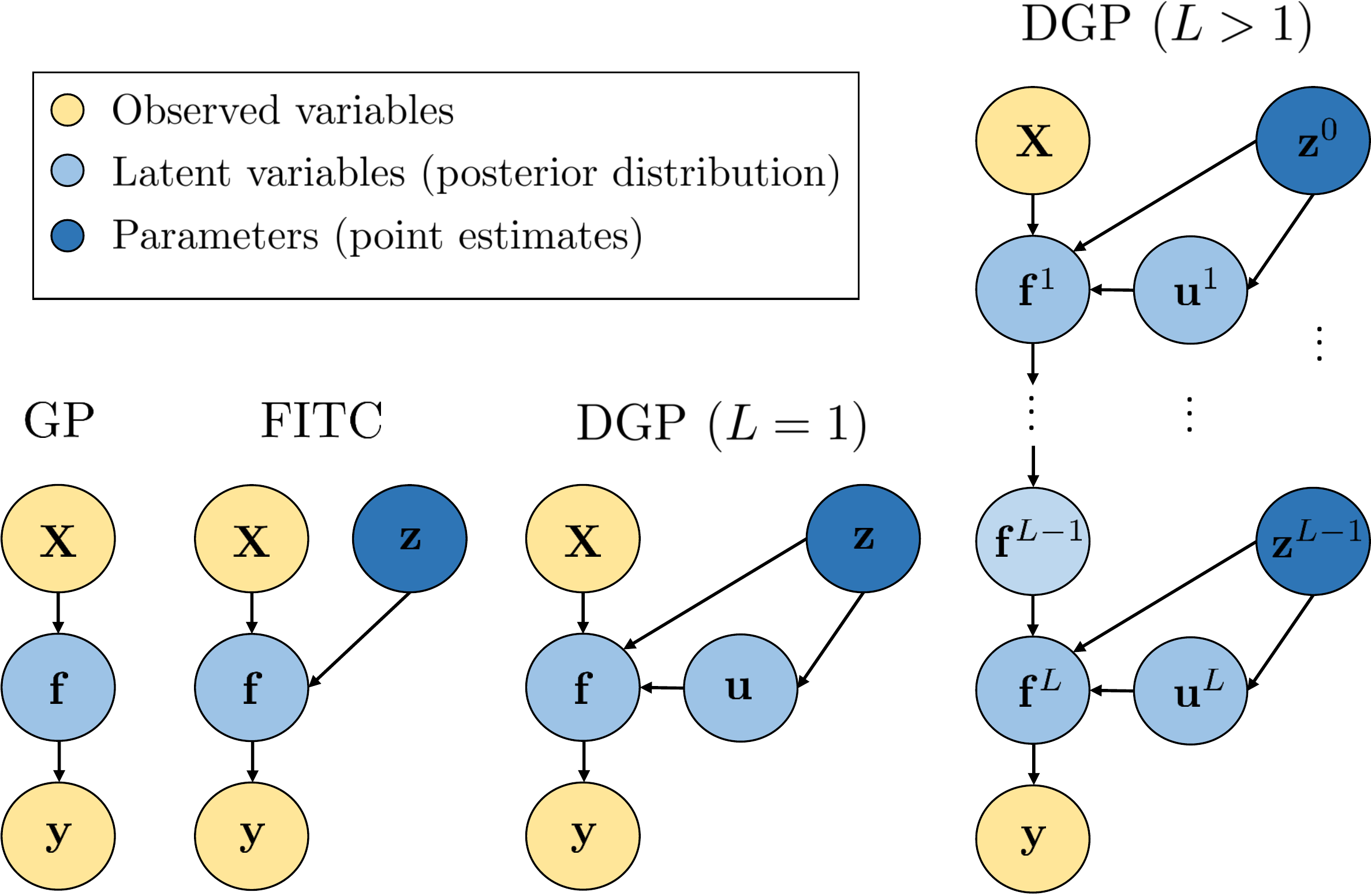}
    \caption{Graphical representation of the four GP-based models used in this work. The color indicates whether a variable is observed or must be estimated. In the latter case, the intensity of the color represents the type of estimation: either through a posterior distribution (light), or a point value (dark).}
    \label{fig:pgms}
\end{figure}

Interestingly, the second paradigm (exact model plus approximate inference) has proven to translate well to hierarchical concatenations of GPs, yielding the inference process for DGPs that is presented in next section.
This justifies that SVGP will be equivalently referred to as DGP ($L=1$) hereafter.
This is also graphically depicted in Fig. \ref{fig:pgms}. Moreover, as explained before, Fig. \ref{fig:pgms} shows that $\bu$ is integrated out in FITC after the model approximation, whereas it is maintained in DGP ($L=1$), where an (approximate) posterior distribution is calculated for it.
As a general summary, Table \ref{tab:methods} shows the main differences between the four GP-based methods that will be used in this work (standard GP, sparse GP FITC, sparse GP SVGP, and deep GP), which are also represented in Fig. \ref{fig:pgms}.

\begin{table*}[t!]
\centering
\caption{Summary of the main differences between the four GP-based models used in this work. VI = Variational Inference.}
\label{tab:methods}
\begin{tabular}{l|llll}
            & GP  & FITC   &  DGP ($L=1$)  & DGP ($L>1$)  \\ \hline
Model & Exact       & Approx.       & Exact &    Exact      \\
Inference  & Exact & Exact       & Approx. (VI) &    Approx. (VI) \\
Depth   & Shallow   & Shallow & Shallow &   Deep \\
Training cost   &  $\mathcal{O}(n^3)$       & $\mathcal{O}(nm^2)$ & $\mathcal{O}(n_bm^2)$    &   $\mathcal{O}\left(n_{b}m^2 \sum_{l=1}^{L} D^l\right)$ \\
References & \citep{Rasmussen2006} & \citep{snelson2006sparse} & \citep{gpBigData,salimbeni_doubly_2017} & \citep{salimbeni_doubly_2017} 
\end{tabular}
\end{table*}

\subsection{Deep Gaussian Processes} \label{sec:DGP}

In standard (single-layer) GPs, the output of the GP is directly used to model the observed response $\by$.
However, this output could be used to define the input locations of another GP.
If this is repeated $L$ times, we obtain a hierarchy of GPs that is known as a Deep Gaussian Process (DGP) with $L$ layers.
This is analogous to the structure of deep neural networks, which are a cascade of generalized linear models \citep[Chapter 6]{damianou_deep_2015}.
Intuitively, this stacked composition will be able to capture more complex patterns in the training data, recall Fig. \ref{fig:DGP_samples}. See also Fig. \ref{fig:pgms} for a graphical depiction of the DGP model.

DGPs were first introduced in \citep{damianou2013deep}, where the authors performed approximate variational inference analytically.
In order to achieve this tractability, in each layer they define a set of latent variables which end up inducing \emph{independence across layers} in the posterior distribution approximation.
This uncorrelated posterior fails to express the complexity of the deep model, and is not realistic in practice.
To overcome this problem, we present the recent inference procedure of \citep{salimbeni_doubly_2017}, which keeps a strong conditional dependence in the posterior by marginalizing out the aforementioned set of latent variables.
In exchange, analytic tractability is sacrificed.
However, we will see that the structure of the posterior allows one to efficiently sample from it and use Monte Carlo approximations.
As will be justified in Appendix \ref{sec:appendix}, this approach is called \emph{Doubly Stochastic Variational Inference} \citep{salimbeni_doubly_2017}.

DGPs can be used for regression by placing a Gaussian likelihood after the last layer. For notation simplicity, in the sequel the dimensions of the hidden layers will be fixed to one (this can be generalized straightforwardly, see both approaches \citep{damianou2013deep, salimbeni_doubly_2017}). But exact inference in DGPs is intractable (not only computationally expensive as in GPs), as it involves integrating out latent variables that are used as inputs for the next layer (i.e. they appear inside a complex kernel matrix). To overcome this, $m$ inducing points $\bu^l$ at inducing locations $\bz^{l-1}$ are introduced at each layer $l$. Interestingly, we will see that this sparse formulation also makes DGP scale well to large datasets, transferring the scalability of (shallow) sparse GP approximations like SVGP up to hierarchical structures.  

For observed $\{\bX,\by\}$, the regression joint DGP model is

\begin{equation}\label{eq:jointModel}
    \p(\by,\{\bff^l,\bu^l\}_{1}^L)=\p(\by|\bff^L)\prod_{l=1}^L \p(\bff^l|\bu^l;\bff^{l-1},\bz^{l-1})\p(\bu^l;\bz^{l-1}).
\end{equation}
Here, $\bff^0=\bX$, and each factor in the product is the joint distribution over $(\bff^l,\bu^l)$ of a GP in the inputs $(\bff^{l-1},\bz^{l-1})$, but rewritten with the conditional probability given $\bu^l$.
Notice that a semicolon is used to specify the inputs of the GP.
The rightmost plot in Fig. \ref{fig:pgms} shows a graphical representation of the described model.

The Doubly Stochastic Variational Inference for this model is detailed in Appendix \ref{sec:appendix}, see also \citep{salimbeni_doubly_2017}. Basically, assuming that the inducing points are enough to summarize the information contained in the training data, the model log-likelihood can be lower bounded by a quantity (called the Evidence Lower Bound, ELBO) that factorizes across data points. This allows for training in mini-batches, just as in SVGP, which makes DGPs scalable to large datasets. Finally, the prediction of the DGPs for a new test data point is included in Appendix \ref{sec:predictions}.

\subsection{Implementation and practicalities}

Several implementations of DGPs are currently available. In our experiments, we used the code integrated within GPflow (a GP framework built on top of Tensorflow), which is publicly available at \href{https://github.com/ICL-SML/Doubly-Stochastic-DGP}{https://github.com/ICL-SML/Doubly-Stochastic-DGP}.
We also used GPflow to train the standard GP and both sparse GP approaches: FITC and SVGP (equivalently, DGP with $L=1$).
In addition, for the sake of reproducibility, we provide illustrative code and demos in a Jupyter notebook at \href{http://isp.uv.es/dgp/}{http://isp.uv.es/dgp/}. The used data is available upon request.



\section{Experimental results } \label{sec:exp}

The problem of translating radiances to state parameters is challenging because of its intrinsic high nonlinearity and underdetermination. 
We consider two such relevant remote sensing problems which together span both land and ocean application, namely 1) predicting surface level temperature and dew point temperature from infrared sounding data, and 2) predicting chlorophyll content, inorganic suspended matter and coloured dissolved matter from S3-OLCI data. Both problems involve inverting a model using large datasets of different sample size and dimensionality.
In the first problem we compare DGPs with (shallow) standard and sparse GPs, highlighting the benefit of going deep in the GP setting. We also illustrate the predictive power of the DGP as a function of depth and data scale.
The second problem aims at comparing the proposed model to another state-of-the-art method in a challenging real application.
Specifically, we compare the performance of a DGP architecture with that of a state-of-the-art neural network method described in \citep{Hiero17}.

\subsection{Surface temperature and moisture from infrared sounders}

Temperature and water vapour are essential meteorological parameters for weather forecasting studies. Observations from high spectral resolution infrared sounding instruments on board satellites provide unprecedented accuracy of temperature and water vapour profiles. However, it is not trivial to retrieve the full information content from radiation measurements. Accordingly, improved retrieval algorithms are desirable to achieve optimal performance for existing and future infrared sounding instrumentation. The use of MetOp data observations has an important impact on several Numerical Weather prediction (NWP) forecasts. The Infrared Atmospheric Sounding Interferometer (IASI) sensor is implemented on the MetOp satellite series. In particular, IASI collects rich spectral information to derive temperature and moisture~\citep{EUMETSAT-IASI-L1,TOURNIER2002}. EUMETSAT, NOAA, NASA and other operational agencies are continuously developing product processing facilities to obtain L2 products from infrared hyperspectral radiance instruments, such as IASI, AIRS or the upcoming MTG-IRS. 
Nonlinear statistical retrieval methods, and in particular kernel machines and Gaussian processes, have proven useful in retrieval of temperature and dew point temperature (humidity) recently~\citep{CAMPS2012,Laparra15,Laparra17}. Here we explore the use of deep Gaussian processes to retrieve surface temperature and moisture from IASI data.

\subsubsection{Data collection and pre-processing}

The IASI instrument scans the Earth at an altitude of, approximately, 820 kilometers. The instrument measures in the infrared part of the electromagnetic spectrum (specifically between wavenumbers 645~cm$^{-1}$ and 2760~cm$^{-1}$, i.e. at wavelengths from 15.5~$\mu$m to 3.62~$\mu$m) at a horizontal resolution of 12 kilometers over a swath width of, approximately, 2200 kilometers. It obtains a global coverage of the Earth's surface every 12 hours, representing 7 orbits in a sun-synchronous mid-morning orbit, and the data obtained from it are used for meteorological models. Each orbit consists of approximately 92\,000 samples collected at a spatial resolution of $0.5$ degrees. This represents more than one million high dimensionality samples to be processed each day.

Obtaining all the products provided by IASI with classical methods requires an enormous computational load. Each original sample has $8461$ spectral bands, but following previous recommendations~\citep{CAMPS2012} we performed feature selection removing the most noisy bands and keeping $4699$. Then we projected the data into the top 50 principal components to combat the risk of overfitting when working with such a high dimensional space. Each pixel is matched with the temperature and dew point temperature at surface level estimated using the European Center for Medium-Range Weather Forecasts (ECMWF) model.

\subsubsection{Experimental setup}

\begin{figure}[!t]
\centerline{\includegraphics[width=8.4cm]{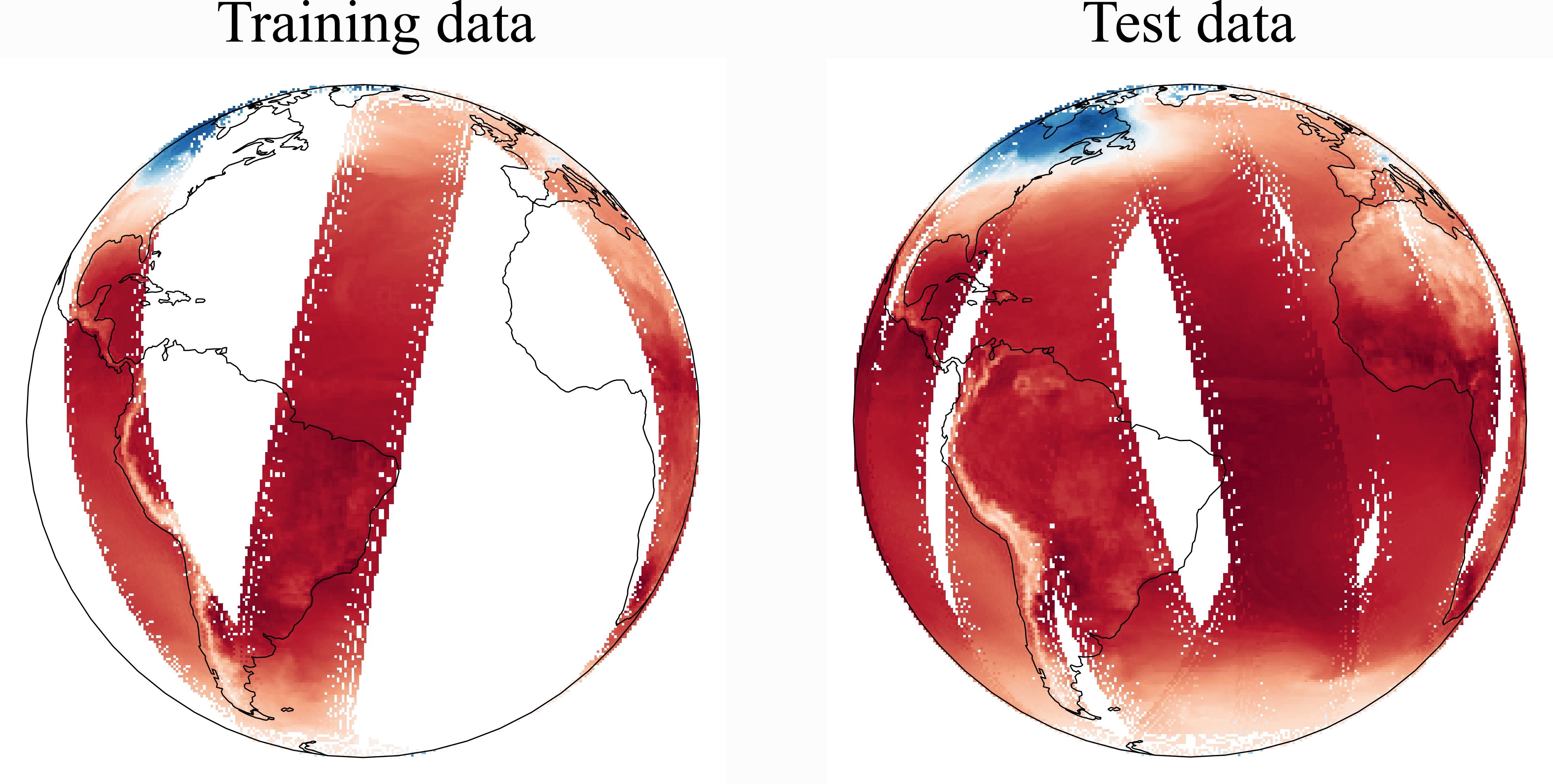}}
\caption{Orbit-wise partition into training and test set used for model comparison when partitioning according to different biomes and climatic zones.} 
\label{fig:cover}
\end{figure}
We employed the data collected in $14$ consecutive orbits within the same day, namely the 1st of January 2013, by the IASI sensor. 
We carried out two different experiments within this application. The first one analyzes how the training data size influences the accuracy in all the GP-related methods, including different depths for the DGP. The second one compares the performance when partitioning the data according to geographical information, such a biome and climate zones. Additionally, it analyzes the quality of the predictive uncertainty.
In the following we refer to these two separate experiments as \textit{Experiment-1} and \textit{Experiment-2} respectively.\\

\noindent \textit{\underline{Experiment-1}:}  In order to analyze the effect of the training data size, we randomly shuffle the data, and select training sets of sizes 10\,000, 50\,000, 140\,000, and 250\,000, and a testing set of 20\,000. 
The root mean squared error (RMSE) that will be reported is the average over five repetitions of the experiment.
The compared models are named as follows:\\

\noindent \textbf{DGP1-4:} DGP described in Section \ref{sec:DGP}
with 1-4 layers and $300$ inducing inputs per layer. The number of hidden units per layer is $5$. 
Recall that DGP1 is equivalent to the sparse GP method SVGP, and the computational cost of DGP is $\mathcal{O}\left(n_{b}m^2(D^1+\dots+D^L)\right)$.

\vspace{0.1cm}

\noindent \textbf{FITC:} Introduced in Section \ref{sec:sparseGP}. Along with SVGP, it is the most popular sparse GP approximation. The RBF kernel is used, and the code is taken from GPflow\footnote{\href{https://github.com/GPflow}{https://github.com/GPflow}}. The cost of training scales like $\mathcal{O}(nm^2)$, and the number inducing points is $300$.

\vspace{0.1cm}

\noindent \textbf{GP-10K:} A standard GP using 10\,000 training points is provided as a baseline. Recall that this is the limit of a standard GP in practice, since it scales like $\mathcal{O}(n^3)$. Again, the RBF kernel and the GPflow library are used. 

\vspace{0.1cm}

\noindent \textit{\underline{Experiment-2}:} Out of the 14 available orbits we choose 11 for test data, and partition it according to: climatic zone, the dominant biome at the location of each data point, latitude, and whether a data point is located at land or at sea. We then selected training data from the remaining 3 orbits (see Fig.~\ref{fig:cover}) and trained one model from each family: A standard, a sparse and a deep Gaussian Process. The models with their sizes of training data were respectively: A standard GP with 10\,000, a FITC with 250\,000 and a 3-layer DGP 250\,000 data points, the data size reflecting the scalability of each training procedure. Comparing the predictions on the test-dataset of $\sim10^6$ points, we also perform an analysis of the provided estimates of predictive uncertainty. 

\begin{figure}[t!]
\centerline{
\includegraphics[width=0.7\textwidth]{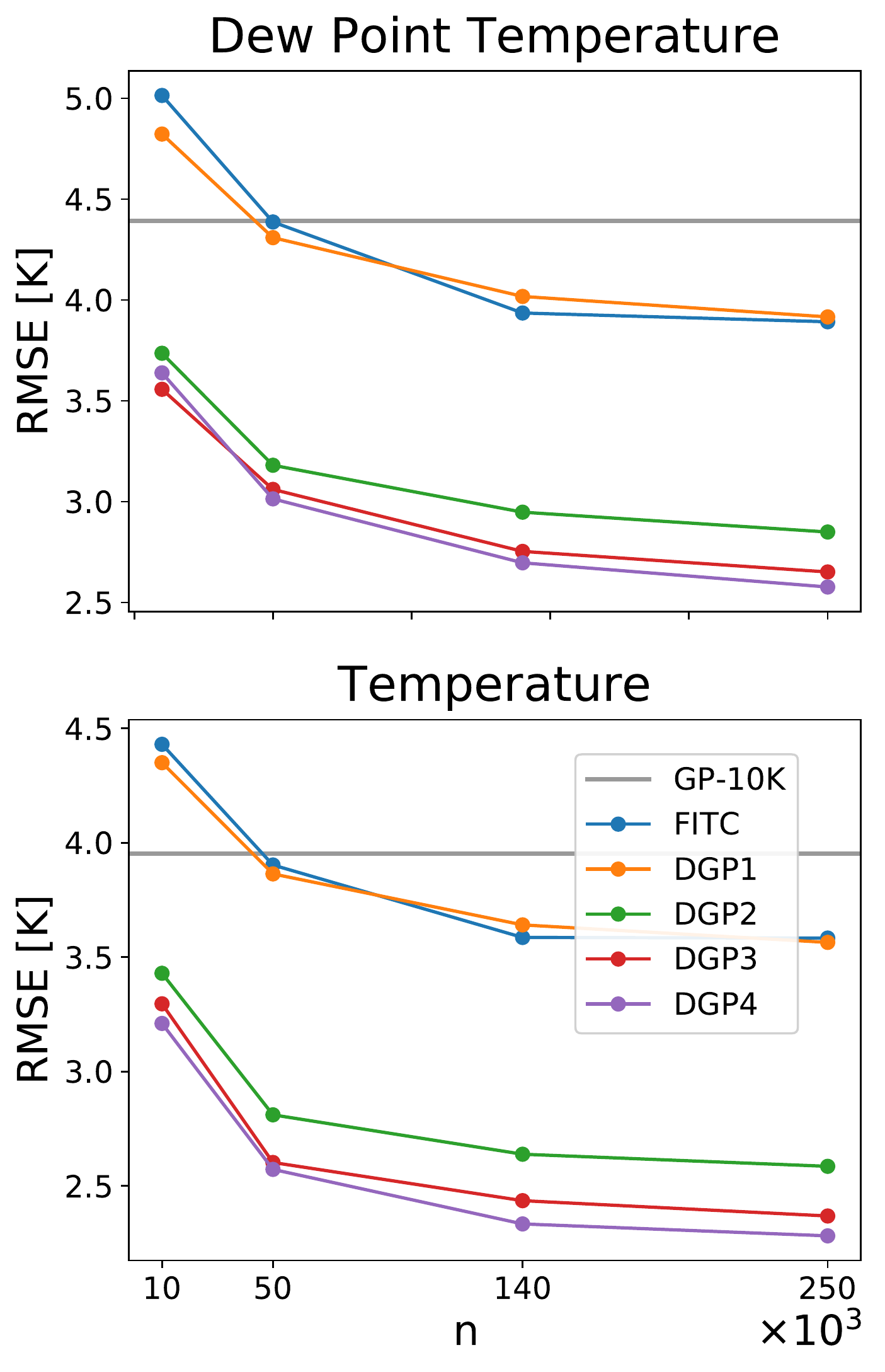}}
\vspace{-0cm}
\caption{Performance of the compared methods as a function of the training set size for the surface dew point temperature (top) and temperature (bottom) variables. The plots share the abscissa. The RMSE of the Deep Gaussian Processes decreases with increasing depth. The DGPs outperform the FITC which performs similarly to the GP-10K.}
\label{fig:rmse} 
\end{figure}

\subsubsection{Experimental results}

The results of \textit{Experiment-1} are summarized in Fig. \ref{fig:rmse}.
We immediately see that there is a clear difference in RMSE between the shallow (GP-10K, FITC, DGP1) and the improved deep models (DGP2-4).
As intuitively expected, the performance of all models increases with additional training data. In this particular problem, it appears that the majority of additional complex structure is learned by going from $10^4$ to $5\times 10^4$ data points. 
As the DGP1 and FITC models are only approximations of the standard GP, it is to be expected that they perform worse when training on the same amount of data as the GP-10K. Nevertheless, when allowed to leverage more data, their fit improves and outperforms the GP-10K. It is not clear which of the two approximations is superior, as it varies with the number of training data. This agrees well with the literature, where this has been shown to depend on the data at hand~\citep{bauer2016}. The fact that single-layer approximations can outperform a standard GP when given enough training data underlines the importance of a model which is able to handle large-scale data. We can see from the results that the DGP both handles large datasets but also allows for higher model complexity and thus a better fit of the data.
From observing the performance of DGPs with different numbers of layers, we can see that DGPs take advantage of their hierarchical structure and achieve lower RMSE with increasing depth. There is a considerable improvement when going from 2 to 3 layers, whereas the effect of going from 3 to 4 layers seems less significant.

\begin{figure}[h!]
\centerline{
\includegraphics[width=0.7\textwidth]{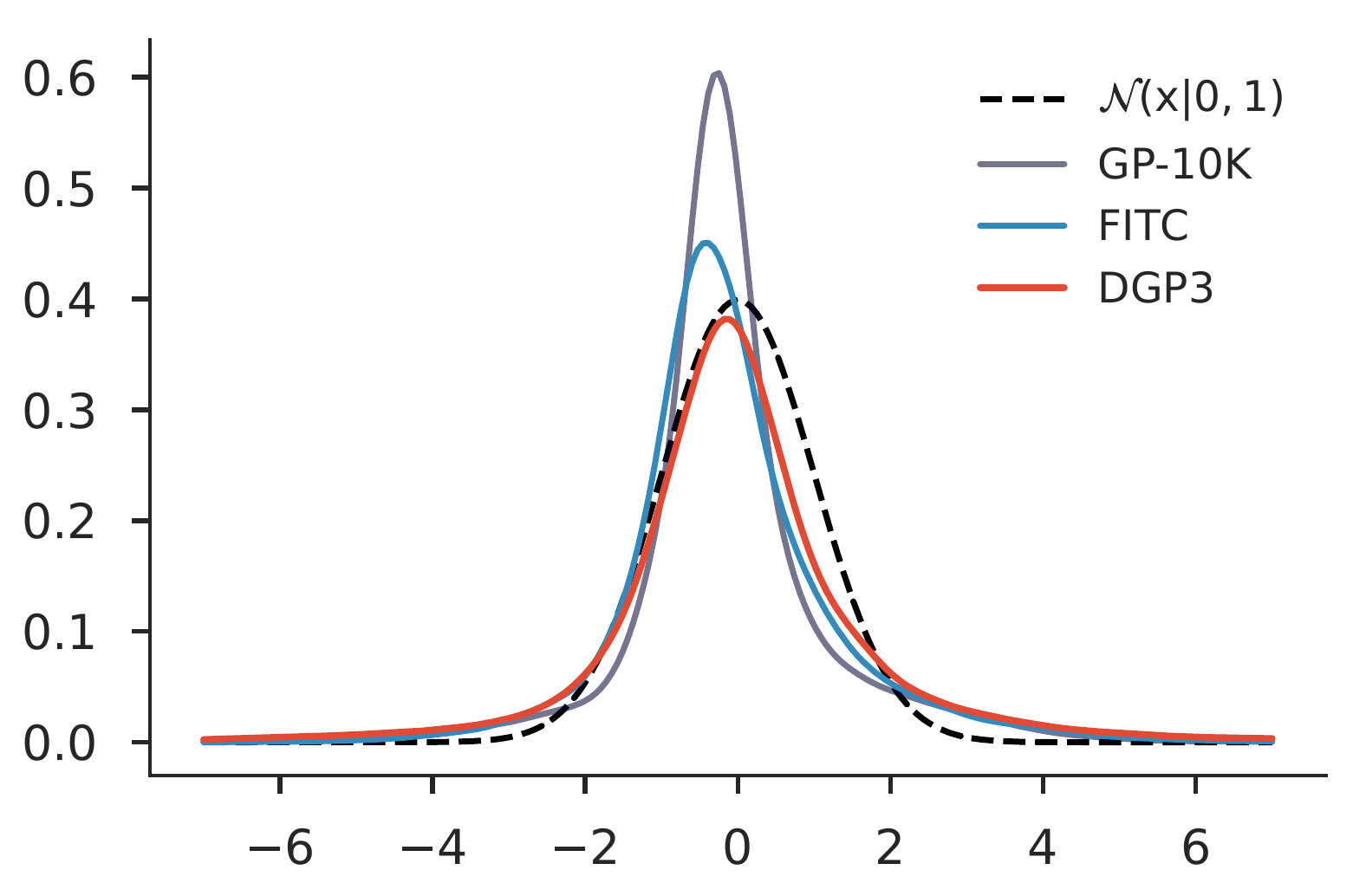}}
\vspace{-0.2cm}
\caption{KDE of residuals normalized by predictive standard deviation, which according to the model should be standard normal distributed. The 3-layer DGP avoids the underestimation seen in the other models, and provides better estimates of predictive uncertainty.}
\label{fig:kde} 
\end{figure}

We now turn to \textit{Experiment-2} for the comparison of the three different GP types, trained according to what their computational cost allows them:
We compare the GP-10K with a FITC and a 3-layer DGP model both trained on 250\,000 data points.
Comparing the predictions on the $\sim$$10^6$ test points (obtained from the 11 orbits shown in Fig.~\ref{fig:cover}) with the ground truth, we can analyze the quality of the predictive uncertainty provided by the models.
Each model provides, for a given test point $y^\ast$, a Gaussian predictive distribution with a mean $\mu(x^\ast)$ and a variance $\sigma^2(x^\ast)$ - see Appendix~\ref{sec:ap_FITC} for the expression for the GP and FITC models, and Appendix \ref{sec:predictions} for the expression for DGPs. Scaling residuals of the predictive mean by the predictive standard deviation we obtain a variable $\zeta^\ast =\frac{\mu(\mathbf{x}^\ast) - y^\ast}{\sigma(\mathbf{x}^\ast)}$ which according to the model should follow a $\mathcal{N}(0,1)$ distribution. Scaling the residuals from prediction on the 11 test orbits in this way, we can make a Kernel Density Estimation (KDE) to analyze their empirical distribution.  The modes of the empirical distributions shown in Fig. \ref{fig:kde} are shifted to the left, indicating a general underestimation in the predictive models. If a model yields too low uncertainties in general (over-confidence), the scaled residuals will become very large and their empirical distributions would have long tails. Conversely, if the model yields too high uncertainties as a rule, the corresponding empirical distribution would be narrowly centered around 0. It can be seen from Fig. \ref{fig:kde} that the scaled residuals of the DGP model follow a $\mathcal{N}(0,1)$ distribution closer than those of the other models, implying that the DGP does the best job of determining the predictive uncertainty correctly. This superior estimate of uncertainty may be due to its higher hierarchical representation capability, accounting for more complex structure in the data.
In practice, this implies improved estimates of how certain the model is about its results when performing parameter retrieval.

\begin{table}[h!]
    \centering
        \caption{\label{tab:2}Mean absolute error (in [K]) and standard error for each model, for predicting surface temperature using different partitions of the test results: land-vs-ocean, climatic zones, per latitude and per biome over land. F and BF are short for Forest and Broadleaf Forest, respectively. Details on climatic and biome classes are given in Appendix~\ref{sec:biome}.}
        \begin{tabular}{l|lll}
        \toprule
        Land/Ocean                  & GP-10K      &     FITC      &  DGP3 \\ \hline
        Land                        & 4.95 (0.02) &   4.96 (0.02) &  \textbf{3.75} (0.02) \\
        Ocean                       & 2.10 (0.01) &   1.96 (0.01) &  \textbf{1.59} (0.01) \\
        \midrule
        Climatic zone               &             &             &                \\ \hline
        Tropical                    & 1.90 (0.03) & \textbf{1.84} (0.03) & 1.88 (0.03) \\
        Arid                        & 5.17 (0.07) & 5.02 (0.06) & \textbf{4.72} (0.07) \\
        Temperate                   & 4.30 (0.18) & 4.60 (0.19) & \textbf{3.54} (0.18) \\
        Cold                        & 6.13 (0.20) & 6.21 (0.20) & \textbf{3.75} (0.16) \\
        Polar                       & 6.81 (0.06) & 6.89 (0.06) & \textbf{4.60} (0.07) \\
        \midrule
        Latitude                    &             &             &                \\ \hline
        $[+40,+60]  $                & 2.09 (0.01) & 2.01 (0.01) & \textbf{1.84} (0.01) \\
        $[+20,+40]  $                & 2.52 (0.02) & 2.22 (0.01) & \textbf{2.20} (0.02) \\
        $[0,+20]   $                & 1.64 (0.01) & \textbf{1.45} (0.01) & 1.58 (0.01) \\
        $[-20, 0]  $                & 2.23 (0.01) & 1.89 (0.01) & \textbf{1.70} (0.01) \\
        $[-40, -20]$                & 3.96 (0.03) & 4.04 (0.03) & \textbf{3.26} (0.03) \\
        $[-60, -40]$                & 5.22 (0.03) & 5.31 (0.03) & \textbf{3.88} (0.02) \\
        \midrule
        Biome                     &              &               &                \\ \hline
        Needleleaf F                & 6.61 (0.11)  &  6.80 (0.11)  &  \textbf{3.38} (0.08) \\
        Evergreen BF                & 1.98 (0.03)  &  1.94 (0.03)  &  \textbf{1.71} (0.02) \\
        Deciduous BF                & 5.02 (0.10)  &  4.83 (0.10)  &  \textbf{2.86} (0.09) \\
        Mixed forest                & 7.45 (0.07)  &  7.48 (0.07)  &  \textbf{5.01} (0.08) \\
        Shrublands                  & 6.44 (0.06)  &  5.66 (0.06)  &  \textbf{4.16} (0.05) \\
        Savannas                    & 2.90 (0.04)  &  2.83 (0.04)  &  \textbf{2.19} (0.03) \\
        Grasslands                  & 6.27 (0.06)  &  6.42 (0.06)  &  \textbf{5.17} (0.07) \\
        Croplands                   & 4.87 (0.05)  &  4.90 (0.05)  &  \textbf{3.44} (0.04) \\
        \midrule \midrule
        Total                       & 5.20 & 5.16 & \textbf{4.17} \\
        \bottomrule
        \end{tabular}
    \label{tab:bigtable}
\end{table}

For the problem of predicting surface temperature, the mean absolute error (MAE) of the trained GP, FITC and DGP3 are shown in Table \ref{tab:bigtable} for different partitions of the test data (and the whole test set in the last row). For the majority of partitions,  DGP3 outperforms the other models, showing that the positive effects of deeper structures are not particular to a type of data, but extend across various meaningful partitions. FITC in turn, being capable of leveraging more data as it does not suffer from the cubic computational cost of the GP, outperforms  GP which is trained on what is widely considered to be its upper limit for the number of training points ($\sim$$10^4$).
In tropical regions of the northern hemisphere, DGP3 performs slightly less accurately than FITC, as seen from the partition into climatic zones and different latitudes. 
Differences in model assumptions and training schemes among machine learning algorithms can cause the models to focus on slightly different parts of the data. It can be concluded, however, that DGPs in general provide much better performance than their shallow counterparts, both due to their ability to leverage large amounts of data and to model more complex data than their shallow counterparts.

\subsection{Ocean color parameters from optical sensors}
Since the first remote sensing images of the ocean were taken, ocean color retrievals have been produced regularly with more or less accuracy, depending on the target parameter, in different regions of the planet and for several water types. The water quality variables reported as able to be estimated by remote sensing are: concentration of inorganic suspended matter (ISM), turbidity, colored dissolved organic matter (CDOM), concentration of chlorophyll-a (Chl-a), occurrence of surface accumulating algal blooms, concentration of phycocyanin, and Secchi depth, e.g.~\citep{Morel1977, Bukata95, Dekker2001}. Research was initially more focused on open ocean or Case-1 waters - where optical properties are determined mainly by the phytoplankton contribution - later with further development of algorithms for more complex or Case-2 waters \citep{Prieur81}. The development and validation of water quality algorithms, many of them empirically developed and implemented using \textit{in situ} data from very specific locations, are the main topic of many of the published investigations. The development of algorithms that do not require extensive \textit{in situ} sampling for training has become an aim in remote sensing of water quality \citep{Kallio183}. For that reason, new databases that combine \textit{in situ} and derived simulated data with radiative transfer models are becoming the training source of semi-analytic and machine learning approaches, like neural networks or Bayesian methods \citep{Doerffer07, Hiero17, Frouin2015}. The experiment carried out here uses one of those recently developed databases, designed within the framework of a European Space Agency (ESA) project called Case 2 eXtreme (C2X), in order to provide a database for the training and validation of a neural net approach \citep{Hiero17}. A subset of this dataset has been already used to test five machine learning approaches, including simple Gaussian processes, for the determination of the three basic ocean colour parameters \citep{Ruescas2018a}.

\subsubsection{Data collection and pre-processing}

Within the framework of the Case 2 eXtreme (C2X) project \citep{Hieronymi2016}, in-water radiative transfer simulations for Sentinel 3-Ocean and Land Instrument (OLCI) were carried out with the commercial software Hydrolight \citep{Mobley2013}. For more detail on the source of the simulations see \citep{Hieronymi2015, Kraseman2016}. In the C2X project, the results of the simulations were grouped into five subcategories: Case 1,  Case 2 Absorbing (C2A),  Case 2 Absorbing-Extreme (C2AX), Case 2 Scattering (C2S) and Case 2 Scattering-Extreme (C2SX), depending on the optical type of water with dominance of absorbing substances (more related to Chl and CDOM) or scattering particles (ISM) in several magnitudes \citep{Hieronymi2016}. Each subcategory consists of 20\,000 individual combinations of concentration of water constituents, inherent optical properties (IOPs), and sun positions. One part of the S3-OLCI simulated dataset is put aside for validation purposes, with more than 4000 spectra per sub-category reserved exclusively for that. The~C2X dataset contains simulations in 21 bands, from which a subset of 11 bands is used here for water quality parameter estimation as in \citep{Hiero17}. This large dataset was used for the training and testing of the S3-OLCI Neural Network Swarm (ONNS) in-water processor \citep{Hiero17}. ONNS is the result of blending various NN algorithms, each optimized for a specific water type, covering the largest possible variability of water properties including oligotrophic and extreme waters. Results from the DGP approach will be compared with the ones achieved by ONNS as part of the validation process.

\subsubsection{Experimental setup}

In the present experiment we have selected all data available for the five categories included in the C2X dataset. In total we have $10^{5}$ records that we use to train and test the DGP models. As already mentioned, 11 out of the 21 S3-OLCI wavebands are selected as inputs, from 400 to 885 nm. The $a_{CDOM}$(440) nm absorption coefficient, using all subgroups (C1, C2A, C2AX, C2S and C2SX), has a range between 0.098 and 20 m$^{-1}$; while the Chl-a content range rises from 0.03 until 200 mg m$^{-3}$; inorganic suspended matter (ISM) ranges from 0.02 to more than 100 g m$^{-3}$. This means that the dataset incorporates a broad range of optical water combinations, making it an effective representation of global ocean and coastal waters including extreme cases.
The purpose of this experiment is to generate the three most popular remote sensing water quality variables (CDOM, Chl-a and ISM) per water category (C1, C2A, C2AX, C2S and C2SX), using DGPs. Other works published on the matter have already used GPs to calculate the three parameters with a subset of the C2X dataset \citep{Ruescas2018a, Ruescas2018b} with promising results. Subsets of the data had to be used in the aforementioned works, as a standard GP cannot leverage data in the order of magnitude presented in the present paper. The DGP model was trained and validated using the same data, $8$$\times$$10^4$ training and $2$$\times$$10^4$ test data points respectvely, as the ONNS \citep{Hiero17}. The results of the ONNS will be used as a source for comparison, that is, we compare our results with state-of-the-art deep learning methods globally accepted in the OC community. 

\subsubsection{Experimental results}

A 3-layer DGP with 500 inducing points and 5 GPs in each hidden layer is trained. Adding more layers was found not to increase performance significantly. This amount of inducing points is frequently used in the GP literature (see e.g.~\citep{shi2019scalable}), and is set to deal with the higher complexity of the C2X dataset.  
Table \ref{T1} shows the comparison of the RMSE between DGP and ONNS dividing the test set by water type, and the total (bottom row). The highlighted results are: compared to ONNS, CDOM results improve in the extreme absorbing and scattering waters, which also affects the total RMSE (DGP 0.115 mg m$^{-3}$ against ONNS 0.202 mg m$^{-3}$). ISM results improve in scattering waters, staying on the same range of error for the other water types, which also translates into almost a factor 3 improvement with the total dataset (DGP 5.296 against ONNS 15.134 g m$^{-3}$). However, the most impressive results are observed for Chl, where more than a factor 3 improvement in the RMSE can be observed for all water cases.

\begin{table}[h!]
\centering
\setlength{\tabcolsep}{1mm}
\caption{Comparison of RMSE between a 3-layer DGP and the ONNS dividing the test set by watertype, and without dividing the test set (bottom row). }\label{T1}
\begin{tabular}{l|l|l|l|l|l|l}
\toprule
\multicolumn{1}{c|}{} & \multicolumn{2}{c|}{\textcolor{black}{\bf CDOM}} & \multicolumn{2}{c|}{\textcolor{black}{\bf ISM}} & \multicolumn{2}{c}{\textcolor{black}{\bf Chl}} \\
\hline
 & DGP & ONNS & DGP & ONNS & DGP & ONNS \\ 
\hline
C1  & 0.0584  & \textbf{0.0174} & 0.1393 &  \textbf{0.0856} & \textbf{2.7913} & 10.2577 \\
C2A & 0.0324  & \textbf{0.0234} & 0.1648 & \textbf{0.116}  & \textbf{2.2126} & 10.5276 \\
C2AX & \textbf{0.2429}  & 0.4356 & 0.2156 & \textbf{0.1429}  & \textbf{2.6765} & 10.3555 \\
C2S & \textbf{0.0200}  & 0.029  & \textbf{0.7784} & 1.4917  & \textbf{2.6668} & 11.2224 \\
C2SX & \textbf{0.0270} & 0.0971 & \textbf{12.476} & 35.689  & \textbf{2.5617} & 16.7635 \\
\hline
Total & \textbf{0.115} & 0.202 & \textbf{5.296} & 15.134 & \textbf{2.594} & 11.914\\
\bottomrule
\end{tabular}
\end{table}

We visualize the behaviour of measured against predicted values in Fig.~\ref{fig:c2x_res}. In this figure the actual values (x-axis) vs. the DGP predictions (y-axis) are compared by variable and water type, in a similar fashion as was done by~\citep{Hiero17} with the ONNS results, with the exception of the non-log scale of our figure. In the following we make references to model predictions in regions of low numerical value which are better appreciated in the log-scale version of the figure located in Appendix~\ref{sec:logfig}.
Summarizing the results by water quality parameter: 
\begin{itemize}
    \item CDOM: {High uncertainties and distribution dispersion in Case 1 and scattering waters (C2S(X)) for very low CDOM values} ($<$0.2 m$^{-1}$). To separate the CDOM from suspended sediments using the absorption signal seems not to be easy. The correlation improves for absorbing waters (C2A(X)) for all values, with good uncertainty ranges for high values in C2A waters, with a gradual increase for the CDOM range higher than 15 m$^{-1}$ in extreme cases.
    \item ISM: Shows almost a perfect correlation for C2S and C2SX, which are the scattering waters where the main component are suspended sediments and non-algal particles. CDOM dominated waters (C2A and C2AX) are not expected to have high non-organic suspended sediment content, which gives less relevance to the more dispersed and less accurate results in these water types. {Some saturation is observed in absorbing waters for very low} values $<$ 0.1 g m$^{-3}$, which is better appreciated in Fig.~\ref{fig:c2x_res_log}, as well as for C1 waters, where dispersion is in general higher; however, it shows lower uncertainty values. This result is in line with the ONNS results in which "the retrieval performance is less skilled if the optical signal of minerals is weak due to low mineral concentrations as is the case in oligotrophic waters (C1)"~\citep{Hiero17}.
    \item Chl: Despite the lower values of uncertainty, there {seems to be some overestimation in the minimum Chl values} (concentrations $<$ 1.0 mg m$^{-3}$) of all five water types. General bias and dispersion is higher in the C2A and C2AX cases. This is an indication of the complexity of the separation of Chl and CDOM for these types of waters. Uncertainty is incremented with high concentration values in all five water types ($>$ 100 mg m$^{-3}$), increasing the dispersion of the data points considerably in C2SX water with Chl values higher than ($>$ 150 mg m$^{-3}$), with a clear underestimation of the parameter. In any case, in nature, cases of extreme ISM and Chl concentration are rare.
\end{itemize}
Considering in the analysis the different water types, Case 1 waters shows quite good results for Chl values ($<$ 100 mg m$^{-3}$), with an increase in the uncertainty and dispersion of the data with higher concentrations. ISM uncertainties are generally low but dispersion and bias are high all through the range. CDOM detection can be problematic and tend to underestimation for values ($<$ 0.5 m$^{-1}$). However, ISM and CDOM are not elements usually found in oligotrophic waters, where Chl is the main contributor to the colour of the water. 
{In C2A and C2AX absorbing waters, Chl values} $<$ 0.5 mg m$^{-3}$ and $>$ 100 mg m$^{-3}$ will be difficult to quantify properly. Lower Chl values would be probably underestimated in C2AX. High Chl concentrations show higher dispersion and uncertainties in both types of water. The ISM distribution is these absorbing waters looks quite good, and uncertainties keep generally low. Dispersion is higher with higher ISM values. CDOM retrievals, however, show a quite good fit to the 1:1 line in C2AX, with an increase of the uncertainty with the increase of CDOM absorption. In C2A waters there is more dispersion around the 1:1 line and more variability in the uncertainty range in values ($<$ 1.0 m$^{-1}$). 
In high scattering waters (C2S and C2SX), there is underestimation in the quantification of high Chl values. CDOM distribution shows cases of over or underestimation depending on the range and water type, with medium to low uncertainties found in low CDOM values. ISM fit to the 1:1 line is very good for both scattering water types, showing C2SX water higher uncertainties in the lower and higher ranges of ISM.

\begin{figure*}[h!]
    \centering
    \includegraphics[width=0.85\textwidth]{figures/_post_review_C2X_result_cmap1_nolog_sidebar.png}
    \caption{Actual values (x-axis) versus values predicted by the DGP (y-axis) of test data, for each of the different water quality variables. The plots are divided by water type, and coloured according to predictive uncertainty: $2*\sqrt{\sigma^2_{DGP}(\mathbf{x})}$.  We see that the predictive uncertainty is generally quite conservative, however it does tend to flag the point of high prediction error with higher predictive uncertainty. This behaviour is preferable compared to the more erratic uncertainty estimates in previous attempts at modeling this data~\citep{Hiero17}.}
    \label{fig:c2x_res}
\end{figure*}

A comparison with the results of~\citep{Hiero17}, can be made, taking into account the several differences between both approaches. On the one hand, the most remarkable fact is that ONNS is a so-called swarm of neural nets designed from and for several water types. The predictions and uncertainties are calculated as the weighted sum of the retrievals of all class-specific NNs. On the other hand, the DGP approach is a single model (pr. output) with three layers and 500 inducing points and 5 GPs per layer. This makes for a more elegant formulation in which there is less choice necessary with respect to model decisions (e.g. 3-4 layers was usually enough to fit the data), and which calculates all uncertainties simultaneously with the retrievals. 
The main success of the experiment is the increase in the accuracy of the Chl quantification for the five different water types. The errors decrease up to a factor 6 in the C2SX water type (see Table \ref{T1}). Uncertainty estimation was found to be a hard problem as previously shown in~\citep{Hiero17}. This is likely due to the fact that the dataset exhibits high variability in regions of both low numeric values and high ones (orders of magnitude from $10^{-2}$ to $10^2$). Nevertheless, the DGP shows some advantages over the ONNS approach: it flags many of the outliers with high predictive uncertainty, and provides more conservative uncertainty estimates than the ONNS which assigns high uncertainty to predictions with low errors as well as vice versa.

\section{Conclusions} \label{sec:con}

We introduced the use of deep GPs and the doubly stochastic variational inference procedure for remote sensing applications involving parameter retrieval and model inversion. The applied deep GP model can  efficiently handle the biggest challenges nowadays: dealing with big data problems while being expressive enough to account for highly nonlinear problems. We successfully illustrated its performance in two scenarios involving optical simulated Sentinel-3 OLCI data and IASI sounding data, and for different data sizes, dimensionality, and distributions of the target bio-geo-physical parameters. 

We showed how DGP benefits from its hierarchical structure and consistently outperforms both full and sparse GPs in all cases and situations on the data at hand. Depth plays a fundamental role but the main increase in performance is achieved when going from shallow to deep Gaussian Processes, i.e. going from 1 to 2 layers. Higher number of layers showed little improvement and a certain risk of overfitting because of model over-parameterization. Importantly, unlike a standard GP, the DGP model is inherently sparse and scales linearly with the training set size. 

We would like to stress that the used DGPs could make a difference in the two applications introduced here, now and in the near future. For instance, neural networks made a revolution in the last decade for the estimation of atmospheric variables from infrared sounding data~\citep{Blackwell05,Blackwell08}. Later, in~\citep{Camps-Valls20121759} we showed that kernel methods can outperform neural networks in these scenarios of high-input and output data dimensionality, but are more computationally costly and memory demanding when bigger datasets are available. With DGPs these shortcomings are remedied: they are more expressive and accurate than standard kernel ridge regression (i.e. one-layer plain GPs), computationally much more efficient, and additionally provide a principled way to derive confidence intervals for the predictions. The problem of estimating temperature and moisture variables was successfully addressed with DGPs, and results were more accurate both over land/ocean, and for different latitudes, climatic zones and biomes. 
Furthermore, the experimental results for prediction of CDOM, ISM and Chl-a showed that it was possible to make a 3-layer DGP outperform a Neural Network based algorithm proposed in the literature. Although uncertainty quantification is difficult, as seen in \citep{Hiero17}, it is an advantage that training a DGP automatically yields uncertainty estimates, avoiding the need to train additional uncertainty neural networks.

The DGP model has demonstrated excellent capabilities in terms of accuracy and scalability, but certainly some future improvements are needed. It does not escape our attention that, as has been shown for deep convolutional neural networks, convolutional models can improve predictions when there is clear spatial structure~\citep{malmgren2017spatial}. Currently there are some efforts in the direction of convolutional GPs~\citep{van2017convolutional}, but performance is still not comparable to a convolutional neural network (CNN).
As shown here and in the literature, DGPs scale very well to large amounts of data, and have been trained on problems with $10^9$ datapoints~\citep{salimbeni_doubly_2017}. As of now, however, feed forward neural networks are still generally faster to train, which is not surprising as the DGP is learning a predictive distribution instead of a single point estimate. Lastly, when it comes to dealing with missing data and mixed data modalities, random forest regression has often been found to be more flexible than other methods. There is interesting work however addressing the missing data problem for GPs~\citep{damianou2015semi}. \\

The more we incorporate machine learning in the pipeline when modeling physical systems, the more important uncertainty estimation and error propagation become. Encoding prior knowledge about input noise into a standard GP in a parameter retrieval setting, it has been shown that improved uncertainty estimation can be achieved~\citep{johnson2019accounting}. The same approach can be imagined with a DGP model, which in the future could additionally improve its uncertainty estimates.

\section*{Acknowledgements}
This work is funded by the European Research Council (ERC) under the ERC-CoG-2014 SEDAL project (grant agreement 647423), the Spanish Ministry of Economy and Competitiveness through projects TIN2015-64210-R, DPI2016-77869-C2-2-R, and the Spanish Excellence Network TEC2016-81900-REDT.
Pablo Morales-\'Alvarez is supported by \textit{La Caixa} Banking Foundation (ID 100010434, Barcelona, Spain) through \textit{La Caixa} Fellowship for Doctoral Studies LCF-BQ-ES17-11600011.
\\ The authors want to thank J. Emmanuel Johnson and Dr. Valero Laparra (Universitat de Val\`encia) for preparing the IASI data; and Martin Hieronymi from Helmholtz-Zentrum Geesthacht and the C2X project for the C2X data set. Hurricane Isabel data used in Fig.~\ref{fig:supertoy} is produced by the Weather Research and Forecast (WRF) model, courtesy of NCAR and the U.S. National Science Foundation (NSF). We give thanks to Dr. David Malmgren-Hansen (DTU, Denmark) for generating Fig.~\ref{fig:cover}.   

\section*{References}
\bibliography{deep,IGARSS18v2}

\begin{appendices}

\section{The Fully Independent Training Conditional (FITC) method}\label{sec:ap_FITC}
Specifically, FITC approximates the model by assuming: i) conditional independence between train and test latent variables $\bff,\bff_*$ given the inducing points $\bu$; and ii) a factorized (fully independent) distribution for $\bff$ given $\bu$. Under these hypothesis, the approximated model for FITC (which replaces the exact $\p(\bff,\bff_*)$) is:
\begin{equation}
    \tilde\p(\bff,\bff_*) = \mathcal{N}\left(\mathbf{0},\left[
    \begin{array}{cc}
       \bQ_{ff} + \mathrm{diag}(\bK_{ff}-\bQ_{ff})  & \bQ_{f*} \\
        \bQ_{*f} & \bK_{**} 
    \end{array}
    \right]
    \right),
\end{equation}
where we abbreviate $\bQ_{ab}=\bK_{au}\bK_{uu}^{-1}\bK_{ub}$. With this approximation, the observation model $\p(\by|\bff,\rho^2)$ can be marginalized
and the new matrix to be inverted is $(\bQ_{ff} + \mathrm{diag}(\bK_{ff}-\bQ_{ff})+\sigma^2\bI)$. Interestingly, this new low-rank-plus-diagonal matrix can be inverted with $\mathcal{O}(nm^2)$ cost by applying the Woodbury matrix identity \citep{Rasmussen2006}.
Finally, the most common practice for the inducing locations $\bZ$ is to estimate them along with the kernel hyperparameters and $\rho^2$ by maximizing the  marginal likelihood   \citep{snelson2006sparse}.

Regarding the predictive distribution, FITC leverages the conditional independence of $\bff_*$ from $\bff$ given $\bu$. Recall that the predictive distribution for a standard GP on a new $\bx_*$ is a Gaussian with mean and covariance given by \citep{Rasmussen2006}:
\begin{align*}
    \mu_{\mathrm{GP}} &= 
    \bK_{*\bff}\left(\bK_{\bff\bff}+\sigma^2\bI\right)^{-1}\by,
    \\
    \sigma^2_{\mathrm{GP}} &=
    \bK_{**}-\bK_{*\bff}\left(\bK_{\bff\bff}+\sigma^2\bI\right)^{-1}\bK_{\bff*}.
\end{align*}
Consequently, the predictive mean and variance for FITC is \citep{snelson2006sparse}:
\begin{align*}
    \mu_{\mathrm{FITC}} &= 
    \bQ_{*\bff}\left(\bQ_{\bff\bff}+ \mathrm{diag}(\bK_{ff}-\bQ_{ff}) +\sigma^2\bI\right)^{-1}\by,
    \\
    \sigma^2_{\mathrm{FITC}} &=
    \bK_{**}-\bQ_{*\bff}\left(\bQ_{\bff\bff}+ \mathrm{diag}(\bK_{ff}-\bQ_{ff}) +\sigma^2\bI\right)^{-1}\bQ_{\bff*}.
\end{align*}

\section{Doubly Stochastic Variational Inference for DGP}\label{sec:appendix}
The approach followed in \citep{salimbeni_doubly_2017} to do inference in DGPs relies on variational inference (VI). The general idea of VI is to transform the problem of posterior distribution computation into an optimization one, by introducing a parametric family of candidate posterior distributions. Moreover, in VI this optimization is solved together with the maximization of the marginal log-likelihood $\log\p(\by)$. More specifically, since the selected family will not usually contain the exact posterior, the target of the optimization will be a lower bound on $\log\p(\by)$.
This is the so-called Evidence Lower Bound (ELBO) \citep{blei2017variational}.

The proposed family of posterior distributions in \citep{salimbeni_doubly_2017} is
\begin{equation}\label{eq:jointPosterior}
    \q(\{\bff^l,\bu^l\}|\{\bz^l,\bm^l,\bS^l\})=\prod_{l=1}^L \p(\bff^l|\bu^l;\bff^{l-1},\bz^{l-1})\q(\bu^l).
\end{equation}
Notice that the first factor is the prior conditional of eq.~\eqref{eq:jointModel}, and keeps correlations between layers. The second is taken Gaussian with mean $\bm^l$ and full covariance $\bS^l$ (which are variational parameters of the parametric family, to be estimated). With this posterior, the ELBO for the marginal log-likelihood $\log\p(\by)$ is then obtained\footnote{The key idea here is that the prior conditionals of eq.~\eqref{eq:jointPosterior} cancel with those of eq.~\eqref{eq:jointModel}. This makes eq.~\eqref{eq:jointPosterior} a very convenient posterior choice.}:
\begin{multline}\label{eq:lowerBound}
    \log\p(\by)=\log\int\frac{\q(\{\bff^l,\bu^l\})}{\q(\{\bff^l,\bu^l\})}\p(\by,\{\bff^l,\bu^l\})\rd\bff^l\rd\bu^l\geq \\
    \sum_{i=1}^n\mathbb{E}_{\q(f^L_i)}[\log\p(y_i|f^L_i)] - \sum_{l=1}^L\mathrm{KL}(\q(\bu^l)||\p(\bu^l;\bz^{l-1})).
\end{multline}
Observe that the second term is tractable, as the KL divergence between Gaussians can be computed in closed form \citep{Rasmussen2006}. However, the expectation in the first term involves the marginals of the posterior at the last layer, $\q(f^L_i)$. Next we see that, whereas this distribution is analytically intractable, it can be sampled efficiently using univariate Gaussians.

Indeed, marginalizing out the inducing points in eq.~\eqref{eq:jointPosterior}, the posterior for the GP layers $\{\bff^l\}$ is
\begin{equation}\label{eq:posteriorF}
    \q(\{\bff^l\})\!=\!\prod_{l=1}^L \q(\bff^l|\bm^l,\bS^l;\bff^{l-1},\bz^{l-1})\!=\!\prod_{l=1}^L \mathcal{N}(\bff^l|\tilde\bmu^l,\tilde\bSigma^l),
\end{equation}
where the vector $\tilde\bmu^l$ is given by $[\tilde\bmu^l]_i=\mu_{\bm^l,\bz^{l-1}}(f^{l-1}_i)$ and the $n\times n$ matrix $\tilde\bSigma^l$ by $[\tilde\bSigma^l]_{ij}=\Sigma_{\bS^l,\bz^{l-1}}(f^{l-1}_i,f^{l-1}_j)$. The explicit expression for the functions $\mu_{\bm^l,\bz^{l-1}}$ and $\Sigma_{\bS^l,\bz^{l-1}}$ can be found in \citep[Eqs. (7-8)]{salimbeni_doubly_2017}. The key point here is to observe that, although the distribution in eq.~\eqref{eq:posteriorF} is fully coupled between layers (and thus the posterior in the last layer is analytically intractable), the $i$-th marginal at each layer $\mathcal{N}(f^l_i|[\tilde\bmu^l]_i,[\tilde\bSigma^l]_{ii})$ only depends on the corresponding $i$-th input of the previous layer. This allows one to recursively sample $\hat f_i^1\to \hat f_i^2\to\dots\to \hat f_i^L$ from all the layers up to the last one by means of just univariate Gaussians. Specifically, $\varepsilon^l_i\sim\mathcal{N}(0,1)$ is first sampled and then for $l=1,\dots,L$:
\begin{equation}\label{eq:sample}
    \hat f_i^l = \mu_{\bm^l,\bz^{l-1}}(\hat f_i^{l-1})+\varepsilon^l_i\cdot\sqrt{\Sigma_{\bS^l,\bz^{l-1}}(\hat f_i^{l-1},\hat f_i^{l-1})}.
\end{equation}

Now, the expectation in the ELBO (recall eq.~\eqref{eq:lowerBound}) can be approximated with a Monte Carlo sample generated with eq.~\eqref{eq:sample}. This provides the first source of stochasticity. Since the ELBO factorizes across data points and the samples can be drawn independently for each point $i$, scalability is achieved through sub-sampling the data in mini-batches. This is the second source of stochasticity, which motivates the naming of this \emph{doubly stochastic} inference scheme.

The ELBO is maximized with respect to the variational parameters $\bm^l,\bS^l$, the inducing locations $\bz^l$, and the kernel and likelihood hyperparameters $\btheta^l$, $\rho^2$ (which, to alleviate the notation, have not been included in the equations). Notice that the complexity to evaluate the ELBO and its gradients is $\mathcal{O}\left(n_{b}m^2(D^1+\dots+D^L)\right)$, where $n_b$ is the size of the mini-batch used, and $D_l$ is the number of hidden units in each layer (which were set to one in this section). As mentioned before, this extends the scalability of the (shallow) sparse GP approximation SVGP \citep{gpBigData} to hierarchical models, including the batching capacity.

\section{Predictions}\label{sec:predictions}
To predict in a new $\bx_*$ in DGPs, eq.~\eqref{eq:sample} is used to sample $S$ times\footnote{This $S$ is related to the first source of stochasticity and, theoretically, the higher the better. In practice, results become stable after a few samples. Here, $S$ was set to 200.} from the posterior up to the $(L-1)$-th layer using the test location as initial input. This yields a set $\{f_*^{L-1}(s)\}_{s=1}^S$ with $S$ samples. Then, the density over $f_*^L$ is given by the Gaussian mixture (recall that all the terms in eq.~\eqref{eq:posteriorF} are Gaussians):
\begin{equation*}
    \q(f_*^L)=\frac{1}{S}\sum_{s=1}^S \q(f_*^L|\bm^L,\bS^L;f_*^{L-1}(s),\bz^{L-1}).
\end{equation*}

\section{Climate zones and biome classification}\label{sec:biome}
The climate zones data were taken from the K\"oppen-Geiger climate classification maps\footnote{See \href{http://koeppen-geiger.vu-wien.ac.at/}{http://koeppen-geiger.vu-wien.ac.at/}.}. The biome zones are aggregations of several classes from the standard International Geosphere-Biosphere Programme (IGBP) biome classification\footnote{For an implementation of the IGBP biome map with 0.05 degree spatial resolution see \href{https://lpdaac.usgs.gov/products/mcd12c1v006/}{https://lpdaac.usgs.gov/products/mcd12c1v006/}.}. The tables below show the IGBP class names and the aggregations performed in this work which are used in Tab.~\ref{tab:bigtable}.

\begin{table}[h!]
    \centering
\begin{tabular}{|l|l|l|}
\hline
\textbf{IGBP name  }               & \textbf{Acronym} \\ \hline
Water                              & WAT            \\ \hline
Evergreen Needleleaf forest        & ENF            \\ \hline
Evergreen Broadleaf forest         & EBF            \\ \hline
Deciduous Needleleaf forest        & DNF            \\ \hline
Deciduous Broadleaf forest         & DBF            \\ \hline
Mixed forest                       & MF             \\ \hline
Closed shrublands                  & CSH            \\ \hline
Open shrublands                    & OSH            \\ \hline
Woody savannas                     & WSA            \\ \hline
Savannas                           & SAV            \\ \hline
Grasslands                         & GRA            \\ \hline
Permanent wetlands                 & WET            \\ \hline
Croplands                          & CRO            \\ \hline
Urban and built-up                 & URB            \\ \hline
Cropland/Natural vegetation mosaic & CVM            \\ \hline
Snow and ice                       & SNO            \\ \hline
Barren or sparsely vegetated       & BSV           \\ \hline \hline
\textbf{Aggregate name} & \textbf{Aggregated classes} \\ \hline
Needle-leaf Forest             & ENF+DNF                \\ \hline
Evergreen Broadleaf Forest     & EBF                  \\ \hline
Decidious Broadleaf Forest     & DBF                  \\ \hline 
Mixed forest                   & MF                  \\ \hline
Shrublands                     & CSH+OSH                \\ \hline
Savannas                       & WSA+SAV                \\ \hline
Herbaceous                     & GRA                 \\ \hline
Cultivated                     & CRO                 \\ \hline
\end{tabular}
    \caption{IGBP biome class names as well as the aggregations of the IGBP classes performed in this work and their corresponding names used in Tab.~\ref{tab:bigtable}.}
    \label{tab:ourlabels}
\end{table}

\newpage
\section{Ocean color results in logscale}\label{sec:logfig}
We include here the results of Fig.~\ref{fig:c2x_res} in log-scale in order to highlight the behaviour of the model when predicting on low numerical values of the parameters.

\begin{figure*}[h!]
    \centering
    \includegraphics[width=0.85\textwidth]{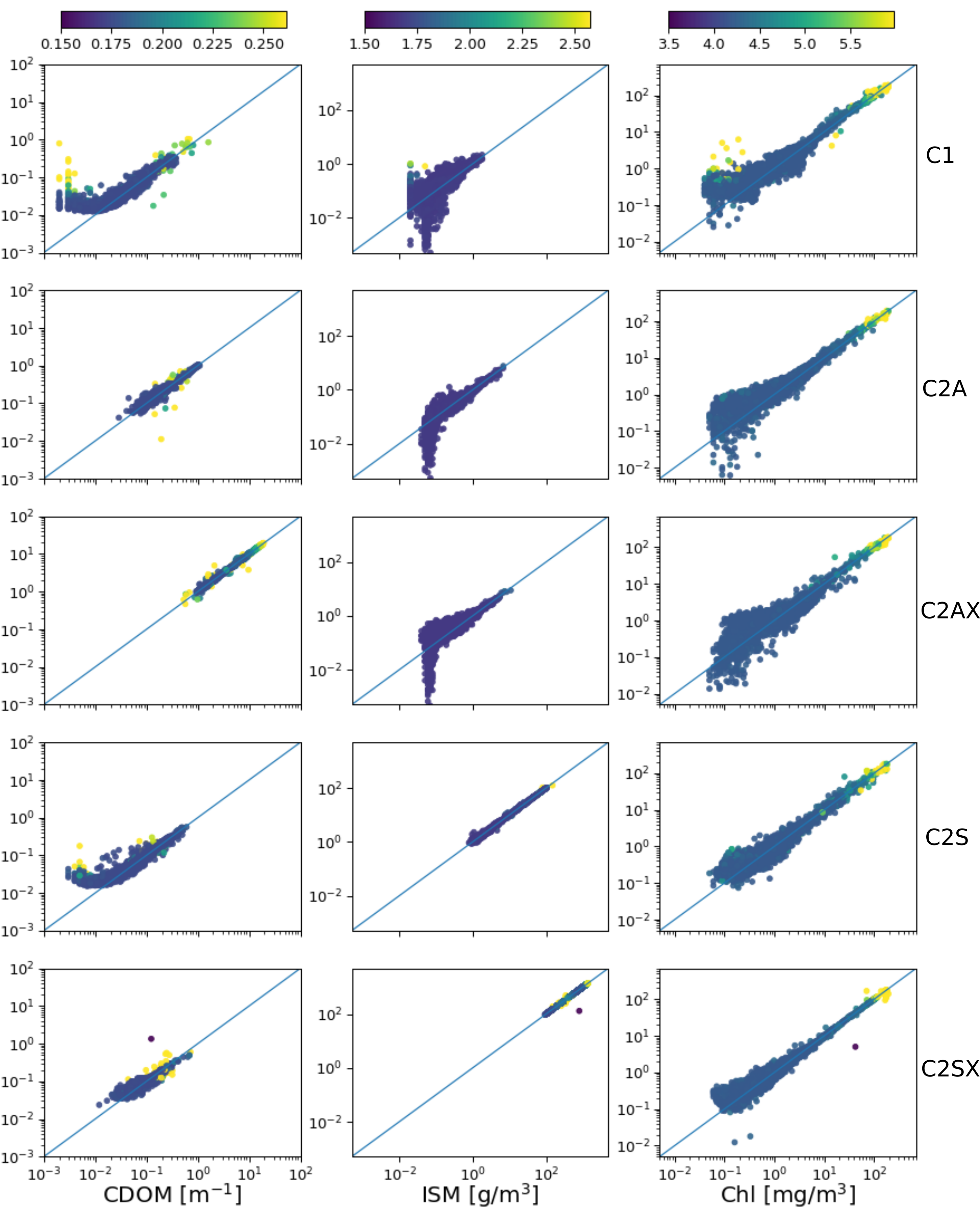}
    \caption{Actual values (x-axis) versus values predicted by the DGP (y-axis) of test data in log scale, for each of the different water quality variables. The plots are divided by water type, and coloured according to predictive uncertainty: $2*\sqrt{\sigma^2_{DGP}(\mathbf{x})}$. 
    }
    \label{fig:c2x_res_log}
\end{figure*}

\end{appendices}
\end{document}